# The terahertz frontier for ultrafast coherent magnetic switching: Terahertz-induced demagnetization of ferromagnets

*Mostafa Shalaby[1], Carlo Vicario[1], and Christoph P. Hauri[1,2]*

[1]*Paul Scherrer Institute, SwissFEL, 5232 Villigen PSI, Switzerland.*

[2]*École Polytechnique Fédérale de Lausanne, 1015 Lausanne, Switzerland.*

*Correspondence to: most.shalaby@gmail.com, carlo.vicario@psi.ch, and christoph.hauri@psi.ch*

*Abstract* The transition frequency between nonthermal coherent magnetic precessions and ultrafast demagnetization is arguably the most sought after answer in magnetism science and technology nowadays. So far, it is believed to be in the terahertz (THz) range. Here, using an ultra-intense low frequency THz bullet, and thin magnetic layers, we report on experimental evidences that fully coherent nonthermal THz magnetic switching may never be reachable in conventional ferromagnetic thin films. At high excitation intensities, while the spins still coherently precess with the THz magnetic field, the deposited THz energy initiates ultrafast demagnetization and ultimately material damage. These series of phenomena are found to take place simultaneously. The reported experiments set fundamental limits and raise questions on the coupling between electronic and magnetic systems and the associated structural dynamics on the ultrafast time scale.

Magnetic data storage relies on the reorientation of magnetic domains which typically occurs on the nanosecond timescale. Future storage technology aims at reorienting the magnetic domains orders of magnitude faster, on the sub-picosecond time scale, by means of an intense laser pulse. The majority of the past studies of ultrafast magnetization dynamics (UMD) on the femtosecond timescale have relied on the indirect electronic excitation via heat deposition [1-4]. Recently, a different opto-magnetic interaction has been shown to be much faster by directly accessing the spin dynamics via the Zeeman torque [5]. The ultimate speed of the coherent spin precession is expected to be at frequencies around $10^{11}$-$10^{13}$ Hz (0.1-10 THz). Reorienting spins at such high carrier frequencies requires enhanced THz field amplitudes (>> MV/cm), which have just become available very recently [6-14]. Recent advances in THz source technology allow for the study of unexplored UMD in the THz frequency range at extremely strong fields [5, 15-16].

So far, light-induced thermal effects as well as the role of the electric field in magnetization dynamics have been neglected at the THz frequencies. This assumption has been partially supported by a recent experiment where magnetic switching has been achieved by in the sub-THz range using magnetic field pulses from relativistic electron beams [17-18]. Using properly shaped THz pulses [19], it was also theoretically shown that the required magnetic field intensity for non-resonant switching increases almost linearly with the triggering frequency [20]. Here, we explore the bottleneck of coherent spin precession by applying an order of magnitude faster (3 THz) stimulus to answer the following fundamental questions: Is magnetic switching in the THz frequency range possible? How do the electronic dynamics affect the coherent spin dynamics? Will the excitation of electronic dynamics eventually limit the speed limit of magnetic switching?

In our present report, we excited thin films of the core magnetic materials (Cobalt, Iron, Nickel) with an ultra-intense THz bullet, targeting switching. We observed that the strong THz transients initiate rich magnetization dynamics on the (sub-) picosecond timescale. The dynamics are dominated by two phenomena. The first is the magnetic field-induced coherent precessions and the second is the energy-induced incoherent demagnetization which can take up to several nanoseconds for thermalization. Both effects are found to be triggered synchronously and on the femtosecond time scale. As the THz excitation intensity is further increased, permanent change of the magnetic properties of the material and surface damage were observed.

As an experimental scheme to measure the THz-induced nonlinear dynamics, we employed collinear THz-pump / magneto-optical Kerr effect (MOKE) probe. The beam configuration at the sample position is shown in Fig. 1A. The external magnetic field **B** (and thus the magnetization vector **M**) was oriented in the sample plane and parallel to the plane of incidence. The reflected optical probe was analyzed to extract the time resolved magnetization dynamics. The measured MOKE rotation contains information on both the in-plane and out-of-plane magnetization dynamics. In order to eliminate any contributions from the nonmagnetic dynamics to the measured signal, we modulated the external magnetic field and locked this modulation frequency to our acquisition system.

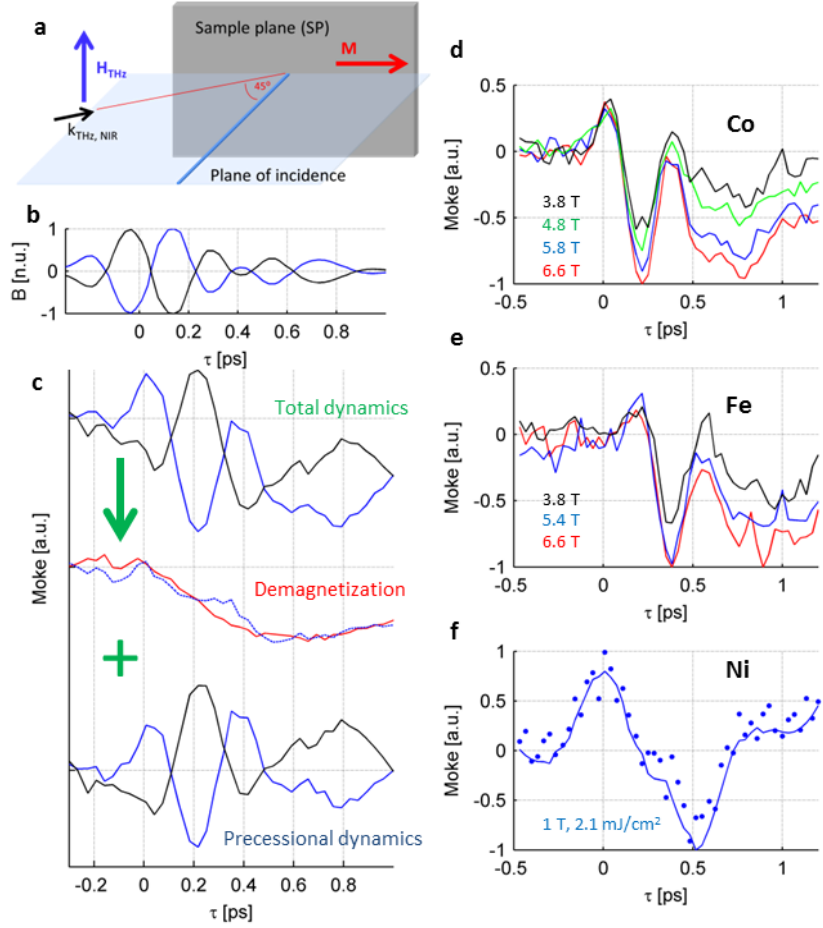

Figure 1. Terahertz excitation of magnetization dynamics. (A) Pump/probe Beam configuration on the sample plane. (B) The triggering THz magnetic fields. (c) The total THz-induced magnetization dynamics (black and blue curves, top plot) under THz magnetic field excitation of 6.6 T. The second plot shows the measured demagnetization components (blue) in comparison with the mathematically extracted one (red). The third plot shows the isolated coherent precessions. The measured dynamics for different levels of fluence in Co (D) and Fe (E). (F) the coherent precessions in Ni measured at very low THz field of 1 T.

Nonthermal coherent magnetic precessions under the application of an external (THz) magnetic field is macroscopically governed by Landau-Lifschitz-Gilbert (LLG) equation:

$$\frac{\partial \mathbf{M}}{\partial t} = -|\gamma|\mathbf{M} \times \mathbf{H}_{\text{eff}} + \frac{\alpha}{M_s}\left(\mathbf{M} \times \frac{\partial \mathbf{M}}{\partial t}\right) \qquad (1)$$

where $\gamma$, $\alpha$ and $M_s$ are the gyromagnetic ratio, Gilbert damping coefficient, and material-dependent saturation magnetization, respectively. $\mathbf{H}_{\text{eff}}$ is the effective field summing the contributions from the internal and external (THz, $\mathbf{H}^{\text{THz}}$) fields. $\mathbf{H}_{\text{eff}}$ is normally aligned to $\mathbf{M}$. An intense enough $\mathbf{H}^{\text{THz}}$ (with a nonzero angle to the direction of $\mathbf{M}$) perturbs this alignment driving time-dependent $\mathbf{M}$ dynamics. The realignment process consists of two stages respectively corresponding to the two terms in the LLG. First, ultrafast precessional dynamics taking place on the THz (stimulus) time scale. Second, damping of **M** towards the new direction of equilibrium $\mathbf{H}_{\text{eff}}$ on a much longer time scale (governed by the material parameters).

In order to maximize the torque and thus nonthermal magnetization dynamics, we preserved $\mathbf{M} \perp |\mathbf{H}^{\text{THz}}|$. The vectorial nature of coherent precessional dynamics suggests that the application of two stimuli with opposite signs $(\pm \mathbf{H}^{\text{THz}})$ leads to opposite torques $\mathbf{M} \times \pm \mathbf{H}_{\text{eff}}$ and thus perfectly reversed temporal dynamics. This opto-magnetic coupling exists even at low THz field strength. However, as the THz field intensities are up-scaled targeting magnetic switching, the amount of THz energy deposited in the magnetic material leads to an increase of temperature and thus demagnetization (similar to magnetic materials excitation with ultrashort optical pulses). Both coherent magnetic precessions and incoherent demagnetization are triggered instantaneously on the THz pulse time scale (Fig. 1B,C for Co at 6.6 T). Our probe (polarization angle normally set to 45°) thus reveals the superposition of the two phenomena as shown in Fig. 1C. By changing the probe polarization angle to 0°, it was possible to significantly increase the sensitivity of the measurement to the in-plane dynamics (demagnetization) as we show in the second plot. The mathematical average of the total dynamics cancels out the coherent precessions and shows only the demagnetization. This average traces well the ultrafast demagnetization measured separately by changing the probe polarization. The coherent precessions can be isolated by removing the demagnetization from the total dynamics as we show in last plot of Fig. 1C.

Figure 1D and 1E show the fluence dependence of the THz-induced dynamics (probe angle = 45°). In the case of Co and Fe, both precessions and demagnetization were obvious at different fluence levels. On one hand, previous studies suggest linear dependence of the precessions on the exciting THz magnetic field (up to a certain level well below the switching threshold). On the other hand, studies of laser-induced demagnetization suggest quadratic dependence of the dynamics on the THz magnetic/electric field. Therefore as the excitation intensity increases, the rate of demagnetization quickly surpasses the precessions thus dominating the measured dynamics. In the case of Ni (Fig. 1F), we could only measure low frequency precessions at low THz magnetic field of 1 T. As we increased the intensities, the demagnetization dominated the measured signal.

In order to simplify the analysis, we looked first at the in-plane demagnetization (by changing the probe polarization angle to 0° where the in-plane dynamics dominated the measured signal). Demagnetization occurs gradually as the coupled THz energy leads to an increase of the temperature of the magnetic material towards the Curie temperature $T_C$ = 1400, 1043, and 627 K for Co, Fe, and Ni, respectively. Neglecting the difference in the THz absorption coefficients between the three materials, Ni is the easiest to demagnetize and Co is the hardest. This qualitative picture is reflected in the quantitative measurements shown in Fig. 2. At a fluence of 89.3 mJ/cm^2, the amount of demagnetization was 14%, 20%, 58%%. The ultrafast demagnetization coincides with the THz pulse and occurs on the scale of 300 fs (likely limited by our relative long sampling optical pulse of with duration of 75 fs). The rate of demagnetization depends on the material and fluence [21]. However, in contrast to the sub-100 fs demagnetization in Ni included by NIR lasers [4], the experimentally obtained demagnetization takes much longer using THz excitation (> 500 fs).

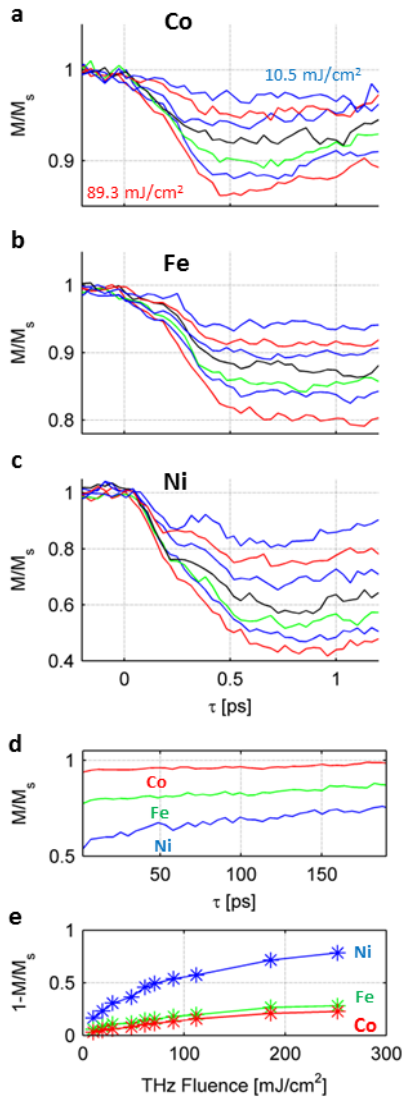

Figure 2. Terahertz-induced demagnetization. (A), (B), (C) The fluence dependence of the ultrafast THz induced demagnetization in Co, Fe, Ni. The excitation levels were 10.5, 19.3, 29.8, 49, 61.3, 71.8, 89.3 mJ/cm$^2$. (D) The delayed recovery of the induced demagnetization under a fluence of 89.3 mJ/cm$^2$. (E) The fluence dependence of the peak demagnetization.

The dynamics described by LLG consist of two parts: the ultrafast precessions and the delayed precessional damping. In the present measurement configuration, the latter was not apparent because of its much lower amplitude in comparison with both the ultrafast precessions and the incoherent demagnetization. However, with a different experimental configuration, we managed to observe it (see supplementary materials). Nevertheless, Fig. 2D shows the delayed demagnetization dynamics were very pronounced extending beyond a delay of 180 ps (limited by our delay line span). The fluence dependence of demagnetization is shown in Fig. 2E. At low fluence, the measurements show nearly linear dependence. Beyond 89.3 mJ/cm$^2$, the magnetic medium was altered mainly in Co and Fe as we discuss next.

As we increased the fluence to 252 mJ/cm$^2$, the demagnetization increased to 80% in the case of Ni. Conversely, Co and Fe started to show material damage. We found that as we increase the excitation fluence even higher, the magnetic materials start to permanently lose their magnetic properties and even show surface damage. Figure 4A shows the demagnetization curves at fluences of 112 mJ/cm$^2$, 186 mJ/cm$^2$, and 252 mJ/cm$^2$. In the case of Co, the initial static magnetization signal (saturation magnetization) is reduced. This reduction is permanent and is increased further when the fluence is further increased. This effect is less manifested in Fe and much less in Ni, which shows a very small reduction. This effect is cumulative over time. Thus the measurements in Fig. 3A are approximate. In Fig. 3B, we measured the reduction of the saturation magnetization with time/THz shot count. Co, Fe, Ni showed reduction by 37 %, 21%, 14% after 150 thousand THz shots. We examined the damage under the scanning electron microscope (SEM). In Fig. 3C, we show the induced damage in Co under a fluence of 252 mJ/cm$^2$.

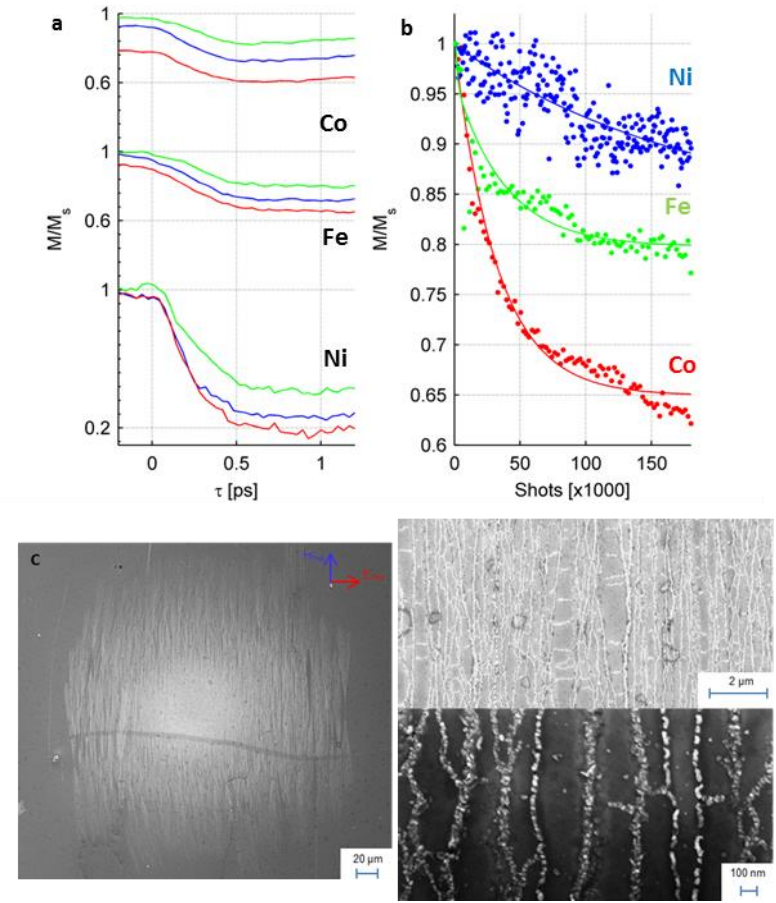

Figure 3. Terahertz-induced damage. (A) The THz induced magnetization at material damaging fluences of (112 mJ/cm$^2$ (green), 186 mJ/cm$^2$ (blue), and 252 mJ/cm$^2$ (red). (B) The degradation in the measured saturation magnetization versus the exposure time for Ni, Fe, and Co at a fixed delay of -10 ps. (C) Scanning microscope images of the damaged Co film at a fluence of 252 mJ/cm$^2$ with a zoom in on the central part of the image.

The shown microscope images shows that the thin film tends to abrade cracking. The break lines tend to be along the normal of the THz electric field. This damage thus contradicts the typical breakdown in dielectrics where the damage occurs along the electric field lines. The physical mechanism of damage is still not well understood. Moreover, what is the contribution of this physical damage (loss of material) to the measured reduction in the saturation magnetization is still unknown. However, it was confirmed during the experiment that the reduction in the saturation magnetization is not mainly due to loss of material or change in the reflectivity of the sample.

In conclusion, we used an ultra-intense THz bullet to trigger ultrafast magnetization dynamics in thin ferromagnetic films where both coherent precessions and incoherent demagnetization were observed. Contrary to the present scientific knowledge, the strong THz transient is found to induce thermal demagnetization and even permanent loss of the magnetic properties and surface damage before magnetic switching occurs. The reported experiments set the frequency limit in coherent magnetic switching and data memory storage speed and raise questions on the already barely-understood coupling between electronic, spin, and lattice systems on the ultrafast time scale.